\documentclass[preprint,showpacs,preprintnumbers]{revtex4}

\usepackage{graphicx}

\begin{document}

\title{Entropy Production of Brownian Macromolecules with Inertia}

\author{Kyung Hyuk Kim}

 \email{kkim@u.washington.edu}

 \affiliation{Department of Physics, University of Washington, Seattle, WA 98195}

\author{Hong Qian}

\affiliation{Department of Applied Mathematics, University of Washington, Seattle, WA 98195}

\date{\today}


\begin{abstract}
We investigate the nonequilibrium steady-state thermodynamics 
of single Brownian macromolecules with inertia under feedback control in isothermal ambient fluid.  With the control being represented by a velocity-dependent external force,  we find such open systems can have a negative entropy production rate and we develop a mesoscopic theory consistent with the second law.  We propose an equilibrium condition and define a class of 
external forces, which includes a transverse Lorentz force, leading to 
equilibrium.  
\end{abstract}






\maketitle

Modern nanotechnology allows the active control of the position and velocity of nanodevices by a feedback system.  The system detects the positions of the nanodevices and differentiates them in manipulating a velocity-dependent external force applied to the nanodevices.  Recently, such a velocity-dependent feedback control (VFC) has been accomplished to reduce the thermal noise of a cantilever in atomic force microscopy (AFM) \cite{Liang} and dynamic force microscopy \cite{Tamayo}.  In \cite{Liang}, a feedback system detects the velocity of the cantilever and reduces its thermal noise by actively changing the direction and magnitude of a force controlling the cantilever according to its motion.  Another VFC experiment has been proposed to control and manipulate a frictional force acting on a small array of particles by limiting 
the terminal velocity of the array with a terminal attractor \cite{Braiman}.  Even though there have been many experiments and theoretical models of VFC on nanodevices, their thermodynamics has been lacking mainly due to the ambiguity in the definition of heat dissipated from the nanodevices \cite{Seki97, Qian-therm}.  This manuscript introduces a thermodynamically-consistent heat \cite{Seki97} and provides the first rigorous theoretical thermodynamic analysis of VFC on nanodevices.  

As models for the above nanodevices, we study single macromolecules  under the VFC in the framework of stochastic dynamics, which has been widely applied to macromolecular processes, e.g., ion channels \cite{Colqu}, motor proteins \cite{Juli97}, biochemical reactions \cite{Qian-biochem}, and nanotechnology.  It is important to study these mesoscopic macromolecular systems \cite{Doi} operating in nonequilibrium steady-state in terms of not only stochastic dynamics but also thermodynamics \cite{Matsuo, Seki97, Rubi98}.  Recently, mesoscopic nonequilibrium thermodynamics of single macromolecules based on Langevin dynamics has been developed in an overdamped regime \cite{Qian-therm, Qian-ent}.  However,  with VFC, inertia plays a fundamental role and its explicit treatment is necessary \cite{Seki97,Kurch98,Lebo}.  Hence, we investigate the relationship among entropy production, detailed balance, and equilibrium in the presence of the inertia.

We discover a novel feature of entropy production rate (EPR).  It is shown to be composed of a positive entropy production rate (PEPR) and an entropy pumping rate (EPuR).  The EPuR indicates how much entropy is pumped out of or into the macromolecule by an external agent manipulating a control force applied to the macromolecules.  The overall entropy production can be negative due to the EPuR term.   This  provides thermodynamic origins of a macromolecular cooling mechanism \cite{Liang, Tamayo, Braiman}.  Furthermore, our approach makes possible the development of macroscopic nonequilibrium steady-state thermodynamics from mesoscopic scale, complementary to an approach with internal degrees of freedom that develops mesoscopic kinetic rules (master equations) from balance equations on hydrodynamic scale \cite{Rubi98}.

Following the general theory of polymer dynamics \cite{Doi}, the macromolecule itself (e.g., a cantilever in the AFM experiment \cite{Liang}) is described by a Hamiltonian, $H(x,y) = \Sigma_i \frac{y_i^2}{2m_i} + U_{int}(x)$, where $x=(x_1,x_2 \cdots x_N)$ and $y=(y_1,y_2 \cdots y_N)$, with $x_i$ and $y_i$ as the 3-D position- and momentum-vectors of the $i$-th hard building block of the macromolecule, respectively.  $U_{int}(x)$ is the internal potential of the macromolecule, e.g, $U_{int}(x) = kx^2/2$ in the AFM experiment, with $k$ a spring constant of the AFM cantilever.  The macromolecule is confined in an isothermal water bath.   
The random collisions between solvent water molecules and the building blocks of the macromolecule are modeled by a Gaussian white noise. This is because the building block is assumed to be much larger than the water molecules and thus the time scales of the two can be separated \cite{Shea96, Shea98}. Using the Einstein summation rule, the Langevin equation for the $i$-th building block of the macromolecule located in phase space at $(X_{i,x}, X_{i,y}, X_{i,z}, Y_{i,x}, Y_{i,y}, Y_{i,z})$ at time $t$ is
\begin{eqnarray}
dX_{i\alpha}/dt&=&\partial_{Y_{i\alpha}} H(X,Y),\nonumber \\
dY_{i\alpha}/dt&=&-\partial_{X_{i\alpha}} H(X,Y) + f_{i\alpha}(Y) \label{stratonovich}\\ 
&&+ {g}_{i\alpha}(X,Y)	+ \Gamma_{i\alpha}^{j\beta} \xi_{j\beta}(t),\nonumber
\end{eqnarray}
where $f_{i\alpha}(Y)$ is an $\alpha$-component frictional force acting on the $i$-th building block by the surrounding water molecules and $g_{i\alpha}(X,Y)$ represents velocity- and position-dependent control by an external agent. In the AFM experiment, the external agent is an electric feedback circuit detecting the motion of the cantilever and manipulating the control force $g$ proportional to its velocity.  In this experiment, $f=-\gamma V$ and  $g=-\alpha V$, with $\gamma$ and $\alpha$ positive constants and $V$ a velocity.  $\Gamma_{i\alpha}^{j\beta} \xi_{j\beta}$ is a fluctuation force caused by collision with water molecules, where $\xi_{j\beta}$ is Gaussian white noise with $\langle \xi_{i\alpha}(t)\xi_{j\beta}(t')\rangle = \delta(t-t')\delta_{\alpha \beta} \delta_{ij}$.  Eq.(\ref{stratonovich}) is studied in terms of its probability distribution, $P(x,y,t)$, using Kramers equation \cite{Gardi}, which is assumed to have a unique stationary state  for our system \cite{Eckmann, Bergmann}.  

Let's consider energy conservation in the Langevin dynamics.  The change of internal energy of the macromolecule, $dH(X_t, Y_t)$, is the same as the work done on the macromolecule by all the external forces; i.e., $dH(X_t, Y_t) = (g+f+\hat{\Gamma} \cdot \xi) \cdot dX$.  The work, $dW(X_t, Y_t)$, done on the macromolecule by the control force $g$,  is $g \cdot dX$.
Then, we may identify the rest of the terms in energy balance as heat, $dQ(X_t,Y_t) \equiv -(f+ \xi \cdot \hat{\Gamma}) \cdot dX$ \cite{Seki97}. This  indicates how much heat is produced and dissipated to the surrounding water heat bath from the macromolecule located at $(X_t,Y_t)$ at time $t$ during time interval $dt$ for a stochastic process.  The energy balance is expressed as $dH =-dQ+dW$.  Note that $(\xi \cdot \hat{\Gamma}) \cdot dX$ denotes $\xi^{i\alpha} \Gamma_{i\alpha}^{j\beta} dX_{j \beta}$.  From the energy balance and Eq.(\ref{stratonovich}), we derive an average heat dissipation rate, 
\begin{eqnarray}
\lefteqn{h_d(t)   \equiv \ \langle dQ/dt \rangle} \nonumber \\
		&=&\int dxdy \{-f \cdot v -\frac{1}{2} Tr( \hat{\Gamma} \hat{\Gamma}^T \hat{M}^{-1})\} P(x,y,t),
\label{dissipation}
\end{eqnarray}
where $\hat{M}_{i\alpha}^{j\beta} \equiv m_i \delta_{ij}$ for any $\alpha, \beta$, with $m_i$ the mass of the $i$-th building block and $v_{i\alpha}\equiv y_{i\alpha}/m_i$ its velocity.  Its detailed derivation will be presented in \cite{Kim}. We note that the stochastic integration is done in Stratonovich sense \cite{Gardi}.

From Eq.(\ref{dissipation}), we find that the frictional dissipation is related to fluctuations.  It implies a fluctuation dissipation relation ($h_d(t=\infty)=0$) in equilibrium and its extension in a nonequilibrium steady state far away from equilibrium.  The average heat dissipation rate can be shown to vanish in equilibrium by substituting the Boltzmann distribution in Eq.(\ref{dissipation}) with the Einstein relation, $\hat{T} \equiv T \hat{I} =  \hat{\Xi}^{-1} \hat{\Gamma} \hat{\Gamma}^T/2 $, where we take $k_B = 1$ and $\hat{\Xi}$ is the frictional coefficient, i.e., $f(y) \equiv -\hat{\Xi}\cdot v$ with $v$ the velocity of the macromolecule, $\hat{I}$ a unit tensor, and $T$ heat bath temperature.

Now we apply the heat dissipation rate, Eq.(\ref{dissipation}), to an entropy balance equation.  Since the above stochastic  process is Markovian, we can introduce Gibb's entropy, $S=-\int P(x,y,t) \ln  P(x,y,t) dxdy$ \cite{Jou99}.  We can connect a  statistical quantity, Gibb's entropy, to  a thermodynamic quantity, heat dissipation rate, using the Einstein relation and find the definite form of the EPR.  We shall find the EPR is composed of two terms: one always positive and the other whose sign depends on the external control force $g(x,y)$.    Using Kramers equation, the entropy balance equation is derived as
\begin{equation}
\frac{dS(t)}{dt} = e_{p+}(t)+e_{pu}(t)-\frac{h_d(t)}{T},
\label{dsdt}
\end{equation}
with 
\begin{eqnarray}
&e_{p+}(t)	&\equiv T^{-1} \int \Pi(x,y,t) \cdot J(x,y,t) dxdy \label{ep+},\\
&e_{pu}(t)	&\equiv \int \{\nabla_y \cdot g(x,y)\}P(x,y,t) dxdy \label{ep-},\\
&h_d(t) 	&\equiv \int J(x,y,t) \cdot f(y) dxdy. \label{h(t)}
\end{eqnarray}
We name $e_{p+}$ and $e_{pu}$ positive entropy production rate (PEPR) and entropy pumping rate (EPuR), respectively.  $\Pi$ is a thermodynamic force defined as the sum of frictional force and Onsager's thermodynamic force; i.e., $\Pi \equiv -\hat{\Xi} \cdot (v + T \nabla_y \ln P)$. $J(x,y,t)$ is a thermodynamic flux corresponding to the thermodynamic force $\Pi$ and is  defined by $-(v+ T \nabla_y \ln P)P$, i.e., the sum of the velocity of the macromolecule and a diffusion flow in momentum space.  Note that, as in macroscopic nonequilibrium thermodynamics \cite{Mazur}, PEPR is expressed as a product of thermodynamic force and its corresponding flux.  Note also that in the above derivation we have used a boundary condition that the macromolecule is confined in the heat bath.

\begin{figure}
\includegraphics[width=8cm]{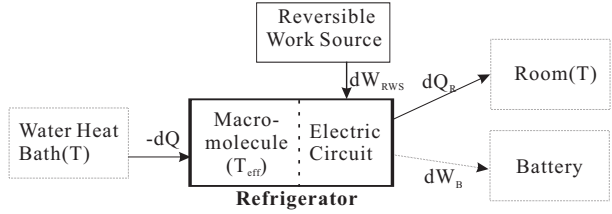}
\caption{\label{fig} A schematic diagram of a 3-D macromolecule immersed in a water heat bath under a control force $\vec{g} = -\alpha \vec{v}$ and frictional force $\vec{f}= -\gamma \vec{v}$ with positive $\alpha$, $\gamma$. The 2nd law of thermodynamics and energy conservation require $dW_B = dW_{RWS}-dQ - dQ_{R} \leq dW_{RWS}$. }
\end{figure}

The PEPR is always non-negative.  This implies the second law of thermodynamics.  To obtain the physical meaning of the PEPR, let an external agent manipulate a control force dependent only on the position of the macromolecule, i.e., $g(x)$.   Then the EPuR vanishes.  The entropy changes due to heat transfer and positive entropy production in nonequilibrium.   In its stationary state, the PEPR is balanced by the heat dissipation rate $h_d$; i.e., the macromolecule constantly dissipates heat to the surrounding water heat bath.  Now, let the external agent manipulate a control force dependent on both the position and the velocity of the macromolecule, i.e., $g(x,y)$.  A new term EPuR appears.  This term is the average of $\nabla_y \cdot g(x,y)$. When $\nabla_y \cdot g(x,y)$ is positive (negative), the distribution of the macromolecule at $y$ in momentum space tends to be dispersed out (contracted in) by the velocity-dependent control force $g$; i.e., the EPuR has a meaning of how much macromolecule's distribution in momentum space is affected by the velocity-dependent control force $g$.  In other words, it describes the amount of entropy  pumped out of (into) the macromolecule to (from) the external agent.  Since the EPuR can be negative, the overall entropy production  can also be negative. A concrete example is the AFM experiment \cite{Liang}.   The stationary distribution of the AFM cantilever is $C Exp[-( y^2/2m+U_{int})/T_{eff}]$, where the \emph{effective} temperature $T_{eff}$ \cite{Vilar01} is $\frac{\gamma}{\alpha+\gamma}T$ and $C$ is a normalization constant. EPR and heat dissipation rate are calculated to be $-3\gamma \alpha / m(\gamma + \alpha)$  and $-3T\gamma \alpha / m(\gamma + \alpha)$, respectively \cite{NegativeEPR}.  In other words, heat flows from water to the macromolecule constantly! The average kinetic energy of the macromolecule is smaller than that of the surrounding water molecules since the control force acts \emph{like} a frictional force on the macromolecule. The kinetic energy is transferred from the water heat bath to the macromolecule.  The macromolecule releases the transferred energy to the external agent.  Let the external agent be an electric circuit connected to a charging battery with infinite storage.  All the transferred energy to the electric circuit can be stored in the battery.  Now, the second law of thermodynamics seems to be violated. This violation stems from the calculation of the entropy of a \emph{portion} of a whole system; the electric circuit and  the macromolecule must be viewed as one \emph{whole} system since they are strongly coupled by a feedback system that detects the velocity of the macromolecule and requires the control force be proportional to the velocity (See Fig.(\ref{fig})).  This combined system  acts like a refrigerator: the macromolecule and the electric circuit act as cooled air  and an engine part of the refrigerator, respectively. If the charging battery is disconnected from the electric circuit, the refrigerator takes heat ($-dQ$) away from the water heat bath and dissipates it outside the electric circuit while also dissipating the work ($dW_{RWS}$) done on the refrigerator, i.e., the work needed to run the electric circuit by detecting the velocity of the macromolecule and requiring the velocity-dependent force.  With the refrigerator connected to the charging battery, a portion of the work ($dW_{RWS}$) done on the refrigerator and a portion of the heat ($-dQ$) transferred from the water is stored in the battery.  This stored energy ($dW_{B}$) must be less than the work done on the refrigerator to satisfy the 2nd law of thermodynamics.

We propose that zero PEPR is equivalent to equilibrium; i.e., when a macromolecular system stops producing entropy, a control force stops working on the system on average and it becomes equilibrated.  This  can be seen from the fact that the equilibrium properties, such as no flux $J=0$ and no heat dissipation $h_d=0$, are satisfied (see Eq.(\ref{ep+}) and (\ref{h(t)})) when PEPR vanishes.   In addition, we identify a class of the control force $g$ leading to equilibrium and it is shown not to work on the macromolecule not only on average but also \emph{instantaneously} if the control force is independent of the  heat bath temperature.  

Now, let's prove that zero PEPR guarantees Boltzmann distribution.   From Eq.(\ref{ep+}), zero PEPR is equivalent to $\nabla_y \ln P = -v/T$.  This means that $P$ is a stationary distribution and is factorized into momentum- and position-dependent parts, where the momentum-dependent part is Gaussian with variance $Tm_i$; i.e., $P_{ss}(x,y) = P_x(x) P_y(y)$, where $P_y(y)=Exp\{-\Sigma_{i=1} ^N y_i^2/2Tm_i\}$.  Plugging this stationary distribution into Kramers equation, we derive $- \nabla_y \cdot g +v \cdot \{g  -\nabla_x U_{int} - T \nabla_x \ln   P_x\}/T= 0$.  This can be rewritten as $\nabla_y \cdot [\{ g(x,y)  -\nabla_x (U_{int}(x)+T \ln  P_x(x))  \}P_y(y)]=0$. Finally, we derive the forms of the control force $g$ leading to the stationary distribution, 
\begin{eqnarray}
g(x,y)&=&A_1(x)+A_2(x,y),
\label{g2}
\end{eqnarray}
where $A_1(x) \equiv \nabla_x(U_{int}(x) + T \ln  P_x(x))$ and $A_2(x,y)$ is any solution satisfying $\nabla_y \cdot \{A_2(x,y) P_y(y)\} = 0$. Since the separation of $g(x,y)$ into $A_1$ and $A_2$ is unique \cite{Unique}, we can define $A_1(x)$ as a conservative external force, i.e., $U_{ext}(x) \equiv -U_{int}(x)- T \ln P_x(x) + C^\prime$ where $C^\prime$ is a constant. Therefore, the probability distribution function $P_{ss}(x,y)$ becomes $C Exp[-( \Sigma_{i=1} ^N y_i^2/2m_i + U_{int}(x) +U_{ext}(x) )/T ]$ where $C$ is a normalization constant.

The velocity-dependent external force, $A_2(x,y)$, leads a system, in which a macromolecule is confined in a heat bath,  to equilibrium \cite{quantum}.  A magnetic force (${A_2}_i = q_i v_i \times B(x_i)$) belongs to this class of forces.  This class also includes other kinds of forces such as ${A_2}_i = v_i \times B(x_i) k(v_i^2)$, $v_i \times \nabla_{v_i} h(v_i)$, and $(1- m_i v^2_{iz}/T) \hat{x}_i + (m_i/T) v_{ix} v_{iz} \hat{z_i}$ with arbitrary functions $k$ and $h$.   The first two forces are perpendicular to velocity so the instantaneous work done by these forces is zero, while the last force is not.  Thus, the instantaneous work by external forces is not required to vanish in equilibrium.  \emph{Average} work, however, vanishes, i.e., $E[ v \cdot A_2(x,y)] = \int v \cdot A_2(x,y) P_{eq}(x,y) dxdy = T \int \nabla_y \cdot A_2(x,y) P_{eq}(x,y)dxdy =T e_{pu}=  0$ from Eq.(\ref{ep-}).  In general, instantaneous work by all temperature-independent forces, $A_2$,  vanishes since $A_2$ is perpendicular to velocity from $\nabla_y \cdot A_2 - A_2 \cdot v/T =0$.

Next, we investigate the relationship between zero PEPR and detailed balance \cite{Kampen}.   Detailed balance stems from time reversal property of the  Hamiltonian governing microscopic dynamics \cite{Mazur}, and  is a necessary condition on equilibrium.  However, it has been proven that detailed balance happens to be equivalent to equilibrium in overdamped system \cite{Qian-time}.  What about a  system with inertia under velocity-dependent control?  We find that detailed balance is also equivalent to equilibrium.

Detailed balance in a stationary state is expressed as  
\begin{equation}
P(\vec{x},t|\vec{x}^\prime, t_0)P_{ss}(\vec{x}^\prime) = P(\epsilon \vec{x}^\prime,t|\epsilon \vec{x}, t_0)P_{ss}(\epsilon \vec{x}),
\label{ext-detail}
\end{equation}
where $\vec{x} \equiv (x,y)$ and $\epsilon \vec{x} \equiv (x,-y)$. Here, the form of the stationary distribution, $P_{ss}(x,y)$, is not specified.  From Eq.(\ref{ext-detail}), the linear operator $\mathcal{L}$ of Kramers equation must satisfy 
\begin{equation}
\int \frac{1}{P_{ss}(\vec{ x})} f_1(\epsilon \vec{x}) \mathcal{L}f_2(\vec{x}) d\vec{x} = \int \frac{1}{P_{ss}( \vec{ x})} f_2(\epsilon \vec{x}) \mathcal{L}f_1(\vec{x}) d\vec{x}
\label{rev1}
\end{equation}
for arbitrary $f_1$ and $f_2$, where $d\vec{x}=dxdy$ and we have used $P_{ss}(\vec{x}) = P_{ss}(\epsilon \vec{x})$ from the integration of Eq.(\ref{ext-detail}) over $\vec{x}$. 
Then, from Eq.(\ref{rev1}), we derive a potential condition \cite{Graham71}
\begin{equation}
\nabla_y \ln  P_{ss}(\vec{x})= (\hat{\Gamma} \hat{\Gamma}^T)^{-1} \cdot \{ -2 \hat{\Xi} \cdot v- g(\epsilon \vec{x})  + g(\vec{x})\}. \label{rev2}
\end{equation}
If $g$ is symmetric under time reversal, Eq.(\ref{rev2}) is simplified as $\nabla_y \ln P_{ss}= -v/T$, i.e., the potential condition becomes equivalent to zero PEPR.  Is the control force $g$ symmetric under time reversal?  Yes. This is because a gravitational force and an electromagnetic force constituting the control force are symmetric under time reversal. The detailed balance Eq.(\ref{ext-detail}), the symmetry relation in operator $\mathcal{L}$ Eq.(\ref{rev1}), and potential condition Eq.(\ref{rev2}) are all equivalent in a Markovian system  with $P_{ss}(\vec{x}) = P_{ss}(\epsilon \vec{x})$.  Therefore, the detailed balance is equivalent to zero PEPR.

In conclusion, (i) we find that EPR can be negative under velocity-dependent feedback control because EPuR can be negative.  (ii) We show that both zero PEPR and detailed balance are equivalent to equilibrium.   These results furnish a thermodynamically consistent mesoscopic theory for nonequilibrium steady-state with negative
entropy pumping. (iii) We identify  a class of external forces $g(x,y)$ that lead to equilibrium and do not work on the macromolecule not only on average but also instantaneously, when this force is independent of the temperature of water heat bath.

\begin{acknowledgments}

The authors thank M. den Nijs, C. Jarzynski, J.M. Rub\'{\i} and  B. Tiburzi for useful discussions.  This work was supported by the Royalty Research Fund of the University of Washington.

\end{acknowledgments}

\bibliography{vcontrol}

\begin{thebibliography}{29}
\bibitem[{Neg()}]{NegativeEPR}
\bibinfo{note}{EPuR and PEPR for the stationary state can be calculated to be
  $-3 \alpha/m$ and $3\alpha^2/m(\gamma+\alpha)$ from Eq.(\ref{ep+}) and
  (\ref{ep-}). Therefore, EPuR+PEPR = $-3\gamma \alpha / m(\gamma + \alpha)$.}
\bibitem[{Uni()}]{Unique}
\bibinfo{note}{If the separation is not unique, there exists a different
  separation $g=A_1^\prime(x) + A_2^\prime(x,y)/P_y(y)$ where $A_2^\prime$ is
  any solution of $\nabla_y \cdot A_2^\prime=0$ and $A_2^\prime \neq A_2$. By
  subtracting these two different forms, we have $(A_1(x) - A_1^\prime(x)) P_y
  = -A_2 + A_2^\prime$. $A_2 - A_2^\prime$ satisfies $\nabla_y \cdot (A_2 -
  A_2^\prime)=0$. Therefore $\nabla_y \cdot \{(A_1(x) - A_1^\prime(x)) P_y\} =
  0$ should hold. $A_1(x)- A_1^\prime(x)$ must be zero.}
\bibitem[{qua()}]{quantum}
\bibinfo{note}{Under a different boundary condition, such as a water heat bath
  connected to particle (macromolecule) reservoirs, the control force $g(x,y)$
  may lead a system to nonequilibrium (e.g., a classical quantum Hall system).}
\end{thebibliography}
\end{document}